\documentclass[aps,prl,twocolumn,superscriptaddress,10pt]{revtex4-1}
\usepackage{CJK}
\usepackage{chemfig}
\usepackage{amsmath}
\usepackage{amssymb}
\usepackage{color}
\usepackage{times}
\usepackage{epsf}
\usepackage{graphicx}
\usepackage{epsfig}
\usepackage{epstopdf}
\usepackage{dcolumn}
\usepackage{bm}
\usepackage{float}
\usepackage{url}
\usepackage{CJK}
\usepackage{ulem}
\usepackage{booktabs}
\usepackage{tabularx}
\usepackage{multirow}
\usepackage{siunitx}
\usepackage{{makecell}}
\setcellgapes{3pt}\makegapedcells
\usepackage{array,collcell}

\newcommand\AddLabel[1]{%
  \refstepcounter{equation}
  (\theequation)
}
\newcolumntype{M}{>{\hfil$\displaystyle}X<{$\hfil}}
\newcolumntype{L}{>{\collectcell\AddLabel}r<{\endcollectcell}}

\def\<{\langle}
\def\>{\rangle}
\def\({\left(}
\def\){\right)}
\def\[{\left[}
\def\]{\right]}

\begin{document}

\title{A ubiquitous thermal conductivity formula for liquids, polymer glass and amorphous solids}

\author{Qing Xi}
\affiliation{Center for Phononics and Thermal Energy Science, China-EU Joint Lab for Nanophononics,
School of Physics Science and Engineering, Tongji University,
Shanghai 200092, China}

\author{Jinxin Zhong}
\affiliation{Center for Phononics and Thermal Energy Science, China-EU Joint Lab for Nanophononics,
School of Physics Science and Engineering, Tongji University,
Shanghai 200092, China}

\author{Jixiong He}
\affiliation{Department of Mechanical and Aerospace Engineering, North Carolina State University, Raleigh, NC 27695, USA}

\author{Xiangfan Xu}
\affiliation{Center for Phononics and Thermal Energy Science, China-EU Joint Lab for Nanophononics,
School of Physics Science and Engineering, Tongji University,
Shanghai 200092, China}

\author{Tsuneyoshi Nakayama}
\affiliation{Center for Phononics and Thermal Energy Science, China-EU Joint Lab for Nanophononics, 
School of Physics Science and Engineering, Tongji University, 
Shanghai 200092, China}
\affiliation{Hokkaido University, Sapporo, Hokkaido 060-0826, Japan}

\author{Yuanyuan Wang}
\affiliation{School of Environmental and Materials Engineering, Shanghai Polytechnic University, Shanghai 201209, China}

\author{Jun Liu}
\email{jliu38@ncsu.edu}
\affiliation{Department of Mechanical and Aerospace Engineering, North Carolina State University, Raleigh, NC 27695, USA}

\author{Jun Zhou}
\email{zhoujunzhou@tongji.edu.cn \\ current address: School of Physics and Technology, Nanjing Normal University, Nanjing 210046
, China}
\affiliation{Center for Phononics and Thermal Energy Science, China-EU Joint Lab for Nanophononics,
School of Physics Science and Engineering, Tongji University,
Shanghai 200092, China}

\author{Baowen Li}
\email{Baowen.Li@Colorado.edu}
\affiliation{Paul M Rady Department of Mechanical Engineering, Department of
  Physics, University of Colorado, Boulder, CO 80305-0427, USA}

\date{\today}

\begin{abstract}

The microscopic mechanism of thermal transport in liquids and amorphous solids has been an outstanding problem for a long time.
There have been several different approaches to explain the thermal conductivities for these systems, for example,
the Bridgman's formula for simple liquids, the concept
of the minimum thermal conductivity for amorphous solids, and the thermal
resistance network model for amorphous polymers. Here, we present a ubiquitous formula to calculate the thermal conductivities of liquids and amorphous solids in a unified way, and compare it with previous ones.  
The calculated thermal conductivities using this formula without fitting parameters are in excellent agreement with the experimental data for these systems.
Our formula is not only providing detailed implications on microscopic mechanisms of heat transfer in these systems, but also solves the discrepancies between existing formulae and experimental data. 

\end{abstract}

\maketitle
%

\begin{table*}[]
\caption{The heat carriers and the formulae for calculating thermal conductivity in different states of materials. }
\begin{tabularx}{12cm}{c c<{\centering} M L c<{\centering}}
\hline
                      &   \textbf{Heat carrier}  &\textbf{Thermal conductivity ($\kappa$)}& \multicolumn{1}{l}{} & \textbf{Ref.}\\
\hline
  Gas    &  molecule   &  \epsilon\eta C_{v} & eq_Gas  & \cite{Loebbook}\\
  Liquid &  vibration of molecule   & 3k_{\rm B}v_{s}n_{\rm mole}^{\frac{2}{3}}  & eq_Liquid & \cite{Bridgman1923,Eyring1941}\\
  Crystal & phonon & \frac{1}{3}C\bar{v}_{s}\Lambda & eq_crystal & \cite{Debyebook,Peierls1929,Kittelbook}\\
  Amorphous solid & vibration of atom/molecule & 1.2k_{\rm B}\bar{v}_{s}n_{\rm atom}^{\frac{2}{3}} & eq_amorphous&\cite{Einstein1911,Cahill1989,Cahill1992,Cahill2017}\\
  Polymer & vibration of molecule & {\left(n_{\rm mole}L \right)}^{\frac{1}{2}} \frac{1}{R_{\rm inter}} & eq_polymer & \cite{zhou2020}\\
\hline
\end{tabularx}
\label{table0}
\end{table*}

Thermal conductivity is a fundamental physical property describing material's capability of heat conduction. It is a macroscopic quantity reflecting the statistical properties of heat carriers such as phonons, electrons, or molecules at microscpic level. 
It has been a challenge to describe this macroscopic quantity in terms of only fundamental
parameters specifying material characteristics. Although many works have been devoted to this
problem, a unified formula for calculating the thermal conductivities
of all these systems is still missing, as the heat carriers are
different in different states of matters.

In this Letter, we first discuss the existing (different) formulae of thermal conductivity for various physical states at high temperatures,  exemplified by
gases, liquids, crystals, amorphous solids and polymers. Then we present a unified formula
for thermal conductivity of these material forms available at high temperatures using only fundamental structural components. 

The thermal conductivity formulae proposed for different forms of materials
are listed in Table I. Equation (1) in Table \ref{table0} is derived from the kinetic theory of
gases (KTG) \cite{Loebbook}, where $\epsilon$ is a pure number, $\eta$
is the viscosity, and $C_{v}$ is the heat capacity at constant
volume. 
Bridgman \cite{Bridgman1923} proposed in 1923 an empirical formula Eq. (2) to calculate
the thermal conductivity of liquids, where $n_{\rm mole}=(\rho/m_{\rm
  mole})$ is the number density of molecules, $\rho$ is the mass
density, and $m_{\rm mole}$ ($M_{\rm mole}=m_{\rm mole} N_{\rm A}$) is
the mass (molar mass) of molecules. $k_{\rm B}$ is the Boltzmann
constant, $v_{s}$ is the sound velocity, $N_{\rm A}$ is the
Avogadro constant, and $3k_{\rm B}$ is the heat capacity per molecule. In
Eq. (2), Bridgman \cite{Bridgman1923} assumed that the neighbouring molecules form a
simple cubic lattice. This model works well for liquids (such
as methanol, carbon tetrachloride, chloroform, and water) and breaks
down for liquids with complicated polyatomic molecules (such as liquid
alkane).
Later on, another formula of thermal conductivity of liquid
$\kappa_{\rm liquid}$ was proposed by Eyring group  \cite{Lin1964} taking into account the parts of gas-like and solid-like
thermal conduction in liquids \cite{Eyring1941}, namely, combining Eqs. (1) and (3) together.

In non-metallic crystalline solids, heat carriers are
phonons - the collective excitation of lattice vibrations 
\cite{Kittelbook}. By treating phonons as gas-like, the thermal conductivity
is given by the kinematic formula Eq. (3) in Table
\ref{table0} \cite{Debyebook,Peierls1929,Kittelbook}, where $C$ is the phonon heat capacity, $\Lambda$ is the mean free path (MFP), and
$\bar{v}_{s}=\frac{1}{3}(v_{sl}+2v_{st})$ is the average sound
velocity where $v_{sl}$ and $v_{st}$ are longitudinal and transverse sound velocities, respectively. 

\begin{figure}[htp]
\centering
\includegraphics[scale=0.5]{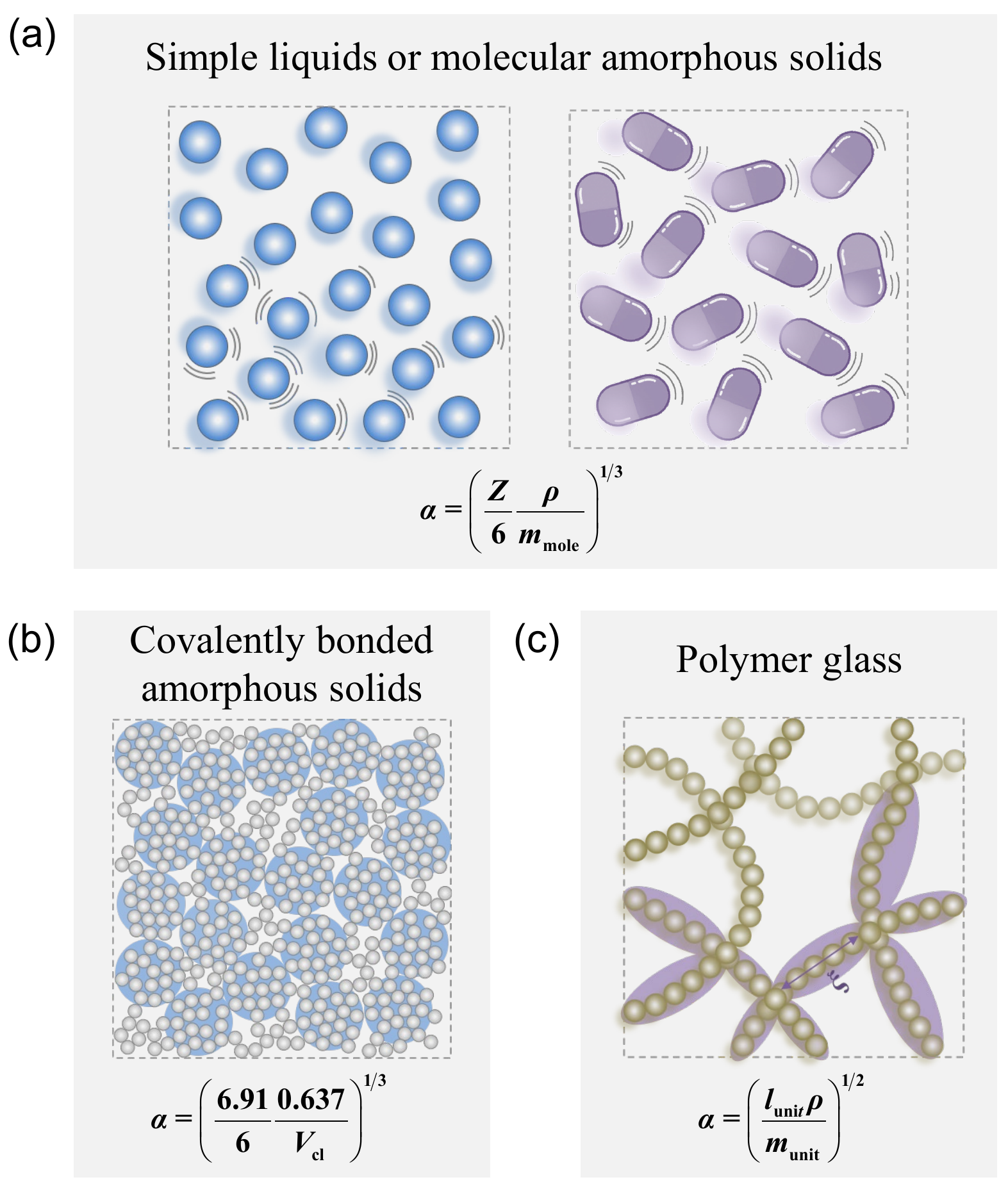}
\caption{(Color online) The schematic illustration of different types of fundamental units transferring heat in liquids and amorphous solids and the prefactor $\alpha$ in Eq.\,(\ref{eq_specify}). (a) The fundamental unit for liquid or molecular amorphous solid is individual molecule. (b) The fundamental unit for  covalent amorphous solid is dense cluster (shaded areas in blue).  $V_{cl}$ is the volume of atomic
 cluster. (c) The fundamental unit for  polymer is a chain segment (shaded areas in purple). 
 }
\label{fig_illustration}
\end{figure}

The derivation of thermal conductivities in amorphous solids ($\kappa_{\rm amor}$) is much less
rigorous because of the ill-defined phonons due to the
lack of lattice periodicity. Kittel \cite{Kittel1949} employed Eq. (3)
to calculate $\kappa_{\rm amor}$ by assuming a nearly constant MFP
($\Lambda_{0}$) which is close to the size of unit cell.
However, Kittel's attempt was not successful in amorphous polymers,
because $\Lambda_{0}$ in polymers should be less than 1$\text \AA$ to recover experimental data from
Eq. (3) \cite{Krevelenbook}. This value is much shorter than the
bond length of molecules. 

Cahill and Pohl \cite{Cahill1989,Cahill1992,Cahill2017} extended this idea to propose the concept of the minimum thermal conductivity (MTC) based on the Einstein's
theory \cite{Einstein1911}. Einstein \cite{Einstein1911} postulated the heat transfer as a random walk of
energy elements between neighbouring localized oscillators with 26 atomic
neighbours. Cahill and Pohl
\cite{Cahill1989,Cahill1992,Cahill2017} modified the Einstein's theory 
by assuming a cluster of atoms as a vibrational ``entity'' where the
size of cluster is one half of the phonon wavelength. 
At high temperatures, the MTC is reduced into a simple form
of Eq. (4), where $n_{\rm
  atom}=(\rho/m_{\rm atom})$ is the number density of atoms, and
$m_{\rm  atom}$ ($M_{\rm atom}=m_{\rm atom} N_{\rm A}$) is the average
mass (molar mass) of atoms.
Equation (4) works well for many inorganic amorphous solids such as vitreous silica (v-SiO$_2$),
whose atomic structure can be described by the continuous random
network (CRN) model \cite{Zallen1998}. However, it significantly
overestimates thermal conductivities of amorphous solids such as
amorphous selenium (a-Se) composed of crown ring Se$_8$ molecules and polymeric chains of Se. Such molecular solids are not described
by the CRN model.

Very recently, a thermal resistance network (TRN) model was proposed by us
to calculate the thermal conductivity of polymers ($\kappa_{\rm
  polymer}$) whose structures can be described by the random coil
model \cite{zhou2020}, see Eq. (5) in Table \ref{table0}, where
$R_{\rm inter}$ is the thermal resistance across contact points
between molecular chains, i.e. the inter-chain resistance. 
$L$ is the length of molecular chains, $n_{\rm mole}L=\rho L/m_{\rm
  mole}=\rho l_{\rm unit}/m_{\rm unit}$.  Note that $l_{\rm unit}$ and $m_{\rm
  unit}$ are length and mass of repeating unit, respectively. 

It is worth pointing out that Eq. (2) of liquids and Eq. (4) of
amorphous solids are analogous because they are proportional to
$n_{\rm mole}^{2/3}$ and $n_{\rm atom}^{2/3}$, respectively, and include the
sound velocity.  One may ask: is there any common physical
mechanism between them? Indeed, Eqs. (2) and (4) can be
rewritten as $\kappa_{\rm liquid} \propto C v_{s}n_{\rm
  mole}^{-\frac{1}{3}}$ and  $\kappa_{\rm amor}\propto C
\bar{v}_{s}n_{\rm atom}^{-\frac{1}{3}}$, respectively, where $C \sim
k_{\rm B}n_{\rm mole}$, and $\sim k_{\rm B}n_{\rm atom}$ at high
temperatures. These two forms are basically the same as Eq. (3) except a
dimensionless prefactor 3 in Eq. (2) and 1.2 in Eq. (4), when choosing
the MFP $\Lambda$ to be on the same order of  magnitude of $n_{\rm
  mole}^{-\frac{1}{3}}$ and $n_{\rm atom}^{-\frac{1}{3}}$.

\begin{table*}[htp]
\caption{Selections of fundamental units and corresponding parameters for different types of materials.}
\begin{tabular}{c c c c c c c c}
\hline
Material type & Fundamental unit & $\tilde{n}$& Typical size of fundamental unit &
 $     \Xi     $ & $\gamma      $ & $D_v   $ & $\delta$\\
\hline
Molecular liquids and solids & nearly-spherical molecules
&$=n_{\rm mole}$ &
$a\times a\times a$& $2     $&   1      & 1$\sim$3 & $= a\approx n_{\rm mole}^{-1/3}$ \\
Molecular liquids and solids & capsule-like molecules & $=n_{\rm
mole}$&
$L\times a\times a$  &   $2     $&    1 & 1 & $\approx a< n_{\rm
  mole}^{-1/3}$\\
Molecular liquids and solids & disk-like molecules &  $=n_{\rm mole}$&
$L\times L\times a$ & $2     $&    1 & 1 & $\approx a< n_{\rm
  mole}^{-1/3}$\\
Covalently bonded amorphous solids & atomic clusters & $\ll n_{\rm atom}$& $\zeta\times\zeta\times\zeta$ & $2     $ & 1 & 2  &
$\approx b< n_{\rm atom}^{-1/3}$\\
Polymer liquids and solids & chain segments & $\gg n_{\rm mole}$& $\xi\times a\times a$  & $4     $ & 1/2  & 1 &$=a^{\prime} \ll n_{\rm mole}^{-1/3}$\\
\hline
\end{tabular}
\label{table2}
\end{table*}

Further comparison of Eq. (5) with Eqs. (2) and (4) makes it clear that the number density
dependence of thermal conductivity in polymers $n_{\rm mole}^{1/2}$ is
different from that in Eqs. (2) and (4) and the sound velocity
dependence vanishes. We notice that the actual value of
$\frac{(Ln_{\rm mole}^{-1/3})^{1/2}}{R_{\rm inter}}$ is of
the same order of magnitude of $k_{\rm B}v_{s}$ because $L$ and
$n_{\rm mole}^{-1/3}$ are typically a few Angstroms and the thermal
resistance $R_{\rm inter}$ is between 0.65-1.6 $\times 10^{10}$ K/W from fitting results
\cite{zhou2020}. Therefore, $\frac{(Ln_{\rm
    mole}^{-1/3})^{1/2}}{R_{\rm inter}}\approx 10^{-20}$ Wm/K is very
close to the value of $k_{\rm B}v_{s}$.

Based on the above analysis, we conjecture that Eqs. (2)-(5) might be expressed as a single formula. The difference for different material lies only in a prefactor.   In the following we shall derive such a formula from the fundamental heat carriers in these materials.

We adopt Einstein's idea \cite{Einstein1911} that
thermal transport is a random walk of heat through the coupling between vibrational modes \cite{simoncelli2019}.
Einstein \cite{Einstein1911} considered that the random walk occurs between an atom and its surrounding neighbours.
The difference from our theory is: we postulate
that molecules or clusters of atoms should work as
individual fundamental units in which the atomic arrangements are ordered or nearly ordered maintaining short-range order.
Then the random walk of heat element through the network formed by these
fundamental units play a key role for the overall thermal conductivity.


Our hypothesis is based on the fact that both liquids and amorphous
solids possess short-range order only \cite{Stachurskibook}. 
The sizes of fundamental units usually range from several Angstroms to several
nanometers as demonstrated by X-ray diffraction measurements \cite{Stachurskibook}. It should be emphasized that
such a length scale of fundamental units is too small to define a local temperature distribution. It is natural that the thermal
equilibrium in each fundamental unit is achieved very fast compared with overall one, not affecting the overall
thermal resistance above room temperature.  
Thus, the thermal transport between neighbouring
fundamental units are governed by the interaction at the contact, for example, via weak van der Waals (VDW)
interaction, H-bond, and chemical bond etc, where these units are
regarded as phase uncorrelated oscillators.

In complex network systems, heat conduction generally depends on the degree of networks (the average coordination number surrounding a constituent element) \cite{networkHeat1,networkHeat2}. We can regard fundamental units as nodes and the
contact between fundamental units as links among nodes in
analogy to the nodes-links-blobs model for complex network systems \cite{nakayama1994}. 
This view makes easy to understand the dominant heat channel in complex network systems.
When a heat current flows across the network, the number of links per unit length
along the direction of heat current is the cubic root of the number density of links. Then we can write the universal formula of thermal conductivity, that valid for a wide range of complex disordered systems, at high temperatures as:
\begin{equation}
\kappa={\left(\frac{Z}{3\Xi}\tilde{n}\right)}^{\frac{1}{3}}h,
\label{eq_main}
\end{equation}
where $Z$ is the degree of network (the average coordination
numbers). 3 is the space dimensionality, $\Xi$ is the number of nodes (fundamental units) sharing one link, and $\tilde{n}$ is the number density of
nodes (fundamental units). $h$ is the heat conductance between blobs, which means that average energy flow across links per unit
time per temperature drop with unit $ JK^{-1}$s$^{-1}$.
The heat conductance can be written as 
\begin{equation}
h=\frac{1}{\gamma}\frac{C_{\rm per} v_{s}}{\delta},
\label{eq_h} 
\end{equation}
where $\gamma$ is the proportion of inter-molecular heat transfer,
$C_{\rm per}$ is the heat capacity per particle whose high-temperature limit is
$(\frac{3}{2}+\frac{D_{v}}{2})k_{\rm B}$ with 3 in the numerator corresponding to the translational degree of freedom and $D_{v}$ the average vibrational degree of
freedom of particle.
The value of $D_{v}$ should be 0 to 3. 
$\delta$ is the length of link, namely, the distance between the neighbouring atoms or
functional group (such as -CH$_{3}$ in paraffin or polymers) at contact points.


Here we apply the formula Eqs.\,(\ref{eq_main}) and (\ref{eq_h}) to calculate thermal conductivity of liquids
and amorphous solids with different fundamental units in
different network structures. Substances are categorized into three
types according to the characteristics of fundamental units, instead of the matter state.
The illustrations in Fig.\,\ref{fig_illustration} show different types of
networks with different fundamental units. In liquids and amorphous solids made
of simple molecules, such as alkane and paraffin, the fundamental units are
individual molecules. In inorganic amorphous solids with covalent bonded atoms
such as v-SiO$_2$, the fundamental unit is a dense cluster of
atoms as marked by blue circles in Fig.\,\ref{fig_illustration}(b).
In polymers made of macromolecules, fundamental units are drawn by purple ellipse in Fig.\,\ref{fig_illustration}(c), namely, the chain segments between adjacent contact points of different chains.
The average segment length is defined by $\xi$.
The fast thermal equilibrium achieved in individual small fundamental unit is
indubitable for all cases above room temperature. 
Thus the thermal transport can be described as a random walk from one
fundamental unit to its touching neighbour units. At low temperatures, these two
processes should be considered simultaneously. 
In this paper, we focus on high temperature thermal conductivity, where the equipartition theorem is valid and quantum effects are negligible. Low temperature cases will be considered elsewhere.

By properly taking the parameters
for different fundamental units,
Eq.\,(\ref{eq_main}) reduces to

\begin{equation}
\kappa=\alpha\frac{1}{\delta}\left(\frac{3}{2}+\frac{D_v}{2}\right) k_B v_s,
\label{eq_specify}
\end{equation}

where the prefactor $\alpha$ differs from state to state as presented in Fig.\,\ref{fig_illustration}. Its physical meaning is given in the
supplementary information.
The fundamental units for different type of liquids
and amorphous solids are summarized in Table \ref{table2}. The parameters $\Xi$,
$\gamma$, $D_v$, $\delta$ are also presented (see details in supplementary information).

\begin{figure*}[htp]
\centering
\includegraphics[scale=0.57]{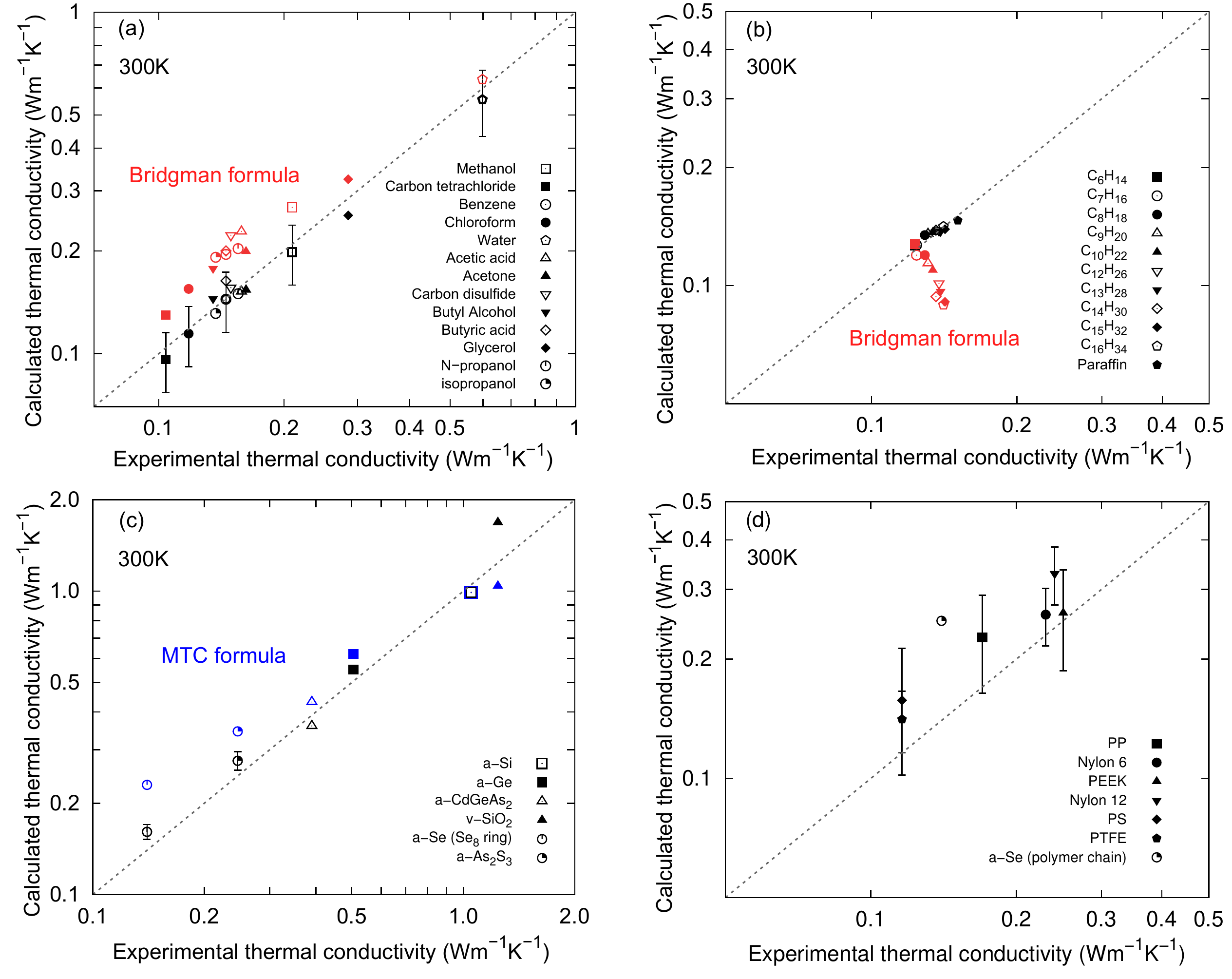}
\caption{(Color online) Comparison of calculated thermal conductivities from  Eq.\,(\ref{eq_specify}) with experimental ones. The diagonal (dotted) line is given to guide for eyes. Symbol lies on this line means calculated thermal conductivity has the same value with experimental one. (a) Calculated thermal conductivities (black symbols) of thirteen
  different liquids at 300 K {\sl vs.} the experimental values. . $D_v$ is chosen
  to be 1$\sim$3 for nearly spherical molecules (methanol, carbon
  tetrachloride, benzene, chloroform, and water) and $D_v=1$ for
  non-spherical molecules. The experimental data of thermal conductivities are taken from
  Refs.\,\cite{Bridgman1923,Yaws1995}. Thermal conductivities calculated from the Bridgman formula (red symbols) are also given for comparison. 
  (b) Calculated thermal conductivities (black symbols) of ten different alkane
  liquids and paraffin at 300 K {\sl vs.} the
  observed values when $D_v=1$. The calculated
  values of $Z$ are listed in supplementary information.
 The experimental thermal conductivities are taken from
  Refs.\,\cite{Yaws1995}. Thermal conductivity calculated according to the Bridgman formula (red symbols) are also given for comparison.
  (c) Calculated thermal conductivities (black symbols) for six different inorganic
  amorphous solids at 300 K {\sl vs.} the observed
  values when $D_v=2$ except for a-As$_{2}$S$_3$ and a-Se, in which case we take As$_{2}$S$_3$ and Se$_8$ molecules as fundamental units with $5<Z<7$ and $D_v=1$. The experimental thermal conductivities are taken from
  Ref.\,\cite{Cahill1992}. Thermal conductivity calculated using the MTC formula (blue symbols) are also given for comparison \cite{Cahill1992}.
  (d) Calculated thermal conductivities (black symbols) for six
  polymers and a-Se composed of 100$\%$ polymeric chains at 300 K {\sl vs.} the experimental values when $D_v=1$. The experimental data are taken from Ref.\,\cite{Mark2009}.}
\label{fig1}
\end{figure*}

We shall point out that the number density of fundamental units is the key parameter in determining the thermal conductivity, as can be seen from Eq.\,(\ref{eq_main}).
When the fundamental unit is the entire small molecule, $\tilde{n}$ can be replace by
$n_{\rm mole}$. When the molecules are large as in polymers, $\tilde{n}$
should be much larger than $n_{\rm mole}$ by choosing chain segments
as fundamental units which contain tens of atoms. Similarly, $\tilde{n}$ should
be the number density of atomic clusters which contains tens of
atoms in covalently bonded amorphous solids.
It needs caution when applying Eq.\,(\ref{eq_specify}) for polymer glasses consisting of extremely long chains, as it is assumed that fundamental units (polymer segments) reach to the thermal equilibrium within the time scale $\tau_{\rm C}$, faster than the time-scale $\tau_{\rm V}$ of the heat conductance between those units. Namely, the situation $\tau_{\rm C}/ \tau_{\rm V}<1$ gives the criterion for the applicability of Eq.\,(\ref{eq_specify}) to polymer glass. 
Since the polymer segments execute coherent vibrations at finite temperature, its characteristic time scale is concerned with the strength of covalent bonding in polymer chain, while the time scale $\tau_{\rm V}$ is governed by the strength of van der Waals force. The ratio can be estimated from the strengths of covalent bonding in polymer chains and van der Waals coupling between chain segments, which is around 100. This means that Eq.\,(\ref{eq_specify})) is applicable for polymer glasses consisting of polymer segments within around 100 times of periodic unit of polymer chains.

Our formula is in excellent
agreement with the experimental data when $1\le D_v\le 3$ for all
substances. We further find that $D_v$ is reduced to 1 when the shape
of molecules is capsule-like or disk-like. This is due to the
shape effect on the liquid relaxation process. In covalently bonded amorphous solids, the
value of $D_v$ is found to be 2.
Possible explanation is the strongly distorted atomic structure at the
edge of fundamental unit where the internal stress shifts from negative value
to positive one \cite{Alexander1998}.

Another major difference of our model from the Bridgman's formula or the MTC one is the intermolecular/interatomic separation
$\delta$. The Bridgman's formula uses $n^{-1/3}_{\rm mole}$ as the intermolecular separation. This is valid only for simple liquids with spherical-like molecules because $a\approx n_{\rm
  mole}^{-1/3}$, where $a$ is approximately equal to the length scale of the occupying space of molecules/atoms. When the molecule is non-spherical, for example,
Se$_8$ is disk-like and acetone is capsule-like, the intermolecular
separation is not the distance between the centers of adjacent
molecules. The distance should be
the separation between contact atoms or functional groups. Therefore,
the value of $\delta$ is approximately $a$ which is obviously much smaller
than $n^{-1/3}_{\rm mole}$.  
In covalently bonded amorphous solids, it is convenient to chose
$\delta$ to be the bond length, $b$, which is smaller than $n^{-1/3}_{\rm
  atom}$.   When the molecule is extremely long in
polymer, $\delta$ is the average value of the size of atoms or functional
groups at the contact points, $a^{\prime}$.

In Fig.\,\ref{fig1}, we compare the calculated thermal conductivities from our formula with experimental values for various liquids, molecular solids, inorganic amorphous solids, and
polymers. The symbols given by black are calculated results obtained from the formula Eq.\,(\ref{eq_specify}), and the diagonal (dotted) lines are used to guide eyes to judge the agreement between calculated and experimental thermal conductivities. 
It shows that our formula agrees very well with the experimental data. For more comprehensive analysis, please refer to the supplementary information. 

Equation\,(\ref{eq_specify}) holds in the temperature regime that fundamental units execute incoherent/independent vibrations respectively. This situation is achieved at high temperatures where the measured thermal conductivities are saturated or show weak temperature dependence. This temperature regime has been universally observed for a wide range of complex disordered materials. We note that Eq.\,(\ref{eq_specify}) recovers the weak temperature dependence of $\kappa$ via the temperature dependences of macroscopic parameters included in Eq.\,(\ref{eq_specify}) such as mass density, distance between neighboring fundamental units, and sound velocity.

We shall emphasize here that thermal conductivities are highly dependent on the detailed atomic/molecular structure of fundamental unit. We further point out that the mechanism of thermal conductivity
of liquids and non-crystalline solids has the same origin. For heat transport, liquids should behave as amorphous solids rather than as fluids, when the oscillation period $1/\omega$ of atoms in liquids is much
shorter than their structure relaxation time $\tau$, namely, $\omega\tau>1$ \cite{Tranchenko2013,Frenkelbook,Dyre2006}.

In summary, by introducing the view of the fundamental units characterizing the types of complex disordered systems such as liquids, amorphous solids, and amorphous polymers, we have proposed a ubiquitous
formula for thermal conductivity that applicable for both liquids and
amorphous solids. 
With properly defined fundamental units, we have demonstrated that the thermal conductivity is
dominated by the thermal resistance between neighbouring fundamental units. This is because the fundamental units reach thermal equilibrium much faster than the heat transfer between units.

Our formula Eq.\,(\ref{eq_main}) not only involving deeper implications compared with the existing well-accepted
empirical/theoretical ones, but also solves the discrepancies between these formulae and experiment measurements.

{\bf Acknowledgments}
This work is supported by the National Key R{\&}D Program of China
(No. 2017YFB0406004), the National Natural Science Foundation of China
(No. 11890703). JH and JL are supported by the National Science Foundation (Award
number CBET-1943813 ) and the Faculty  Research and Professional
Development Fund at North Carolina State University.

\end{document}